\documentclass[runningheads]{llncs}

\usepackage[T1]{fontenc}
\usepackage{float}
\usepackage{graphicx}
\usepackage{amsmath}
\usepackage{amssymb}
\usepackage{colortbl}
\usepackage{array}
\usepackage[table,xcdraw]{xcolor}
\usepackage{colortbl}
\newcommand{\highlight}[2]{\colorbox{#2}{#1}}
\newcommand{\ra}{\textrightarrow}

\begin{document}

\title{Unveiling Sleep Dysregulation in Chronic Fatigue Syndrome with and
  without Fibromyalgia Through Bayesian Networks}

\titlerunning{CFS and CFS+FM Sleep Dysregulation via Bayesian Network}

\author{
  Michal Bechny\inst{1,2} \and
  Marco Scutari\inst{3} \and
  Julia van der Meer\inst{4} \and
  Francesca Faraci\inst{1} \and
  Stéphane Meystre\inst{1} \and
  Benjamin H. Natelson\inst{5} \and
  Akifumi Kishi\inst{6}
}

\authorrunning{M. Bechny et al.}

\institute{
  Institute of Digital Technologies for Personalised Healthcare (MeDiTech), SUPSI \and
  Institute of Computer Science, UNIBE \and
  Istituto Dalle Molle di Studi sull’Intelligenza Artificiale (IDSIA) \and
  Department of Neurology Inselspital, Bern University Hospital and UNIBE \and
  Department of Neurology, Icahn School of Medicine at Mount Sinai \and
  Graduate School of Medicine, The University of Tokyo
}

\maketitle

\begin{abstract}

  Chronic Fatigue Syndrome (CFS) and Fibromyalgia (FM) often co-occur as
  medically unexplained conditions linked to disrupted physiological regulation,
  including altered sleep. Building on the work of Kishi et
  al.~\cite{kishi2011sleep}, who identified differences in sleep-stage
  transitions in women with CFS and CFS+FM, we exploited the same strictly
  controlled clinical cohort using a Bayesian Network (BN) to quantify detailed
  patterns of sleep and its dynamics. Our BN confirmed that sleep transitions
  are best described as a second-order process~\cite{yetton2018quantifying},
  achieving a next-stage predictive accuracy of 70.6\%, validated on two
  independent data sets with domain shifts (60.1--69.8\% accuracy). Notably, we
  demonstrated that sleep dynamics can reveal the actual diagnoses. Our BN
  successfully differentiated healthy, CFS, and CFS+FM individuals, achieving an
  AUROC of 75.4\%. Using interventions, we quantified sleep alterations
  attributable specifically to CFS and CFS+FM, identifying changes in stage
  prevalence, durations, and first- and second-order transitions. These findings
  reveal novel markers for CFS and CFS+FM in early-to-mid-adulthood women,
  offering insights into their physiological mechanisms and supporting their
  clinical differentiation.


\end{abstract}

\section{Introduction}

Chronic Fatigue Syndrome (CFS) and Fibromyalgia (FM) co-occur in up to 70\% of
cases~\cite{brown2007functioning}. These conditions share symptoms such as
disrupted sleep and exhaustion, but have distinct clinical profiles: CFS is
characterised by severe, unexplained fatigue worsened by
exertion~\cite{clayton2015beyond}, whereas FM is defined by widespread
musculoskeletal pain and sensory hypersensitivity~\cite{galvez2020diagnostic}.
Both conditions disproportionately affect females, with prevalence up to four
times higher than in males~\cite{faro2016gender}, and are most commonly reported
in young to middle-aged
adults~\cite{clayton2015beyond,galvez2020diagnostic,natelson2019myalgic}. They
are frequently accompanied by other clinical conditions, including psychiatric
and specific sleep disorders~\cite{unger2004sleep,mariman2013sleep},
complicating the quantification of their underlying effects. Consequently,
clinical reviews of existing---mostly observational---studies often lack
evidence of their systematic impacts on sleep
architecture~\cite{mariman2013sleep}.

The study cohort by Kishi et al.~\cite{kishi2011sleep} minimized confounding
factors and collected polysomnographic (PSG) data from a strictly controlled set
of healthy (H), CFS, and CFS+FM women aged 25–55. Exploratory data analysis
revealed changes in sleep stage durations and proportions, and identified
first-order transitions as potential markers enabling clinical interpretation of
physiological dysregulation in CFS and CFS+FM.

Recent research in individuals with or without sleep disorders showed that
sleep-stage transitions are optimally modelled and analysed as a second-order
process~\cite{schlemmer2015changes,yetton2018quantifying}. Leveraging these
insights and the CFS/FM dataset~\cite{kishi2011sleep}, we (i) \textit{implement
a Bayesian Network (BN) capable of both next-stage prediction and diagnostics},
(ii) \textit{validate the second-order optimality even in a clinical cohort},
and based on that (iii) \textit{identify novel markers for CFS and CFS+FM based
on two-stage transitions,} providing novel insights into their physiology and
supporting their clinical differentiation.

\section{Materials and Methods}

\subsection{Data}

\textbf{Primary Cohort.} The data from \cite{kishi2011sleep} comprises PSG
recordings from 52 women, carefully selected to ensure homogeneity and avoid
confounding. The cohort included 26 healthy controls (H, aged 38 $\pm$ 8 years),
14 individuals with CFS only (aged 37 $\pm$ 9 years), and 12 individuals with
CFS and FM (CFS+FM) (age: 41 $\pm$ 6 years). Rigorous exclusion criteria were
applied, including the presence of clinically evident sleep disorders or other
psychiatric conditions. Subjects also refrained from alcohol, caffeine and
strenuous activities before the study, and menstruating individuals were
evaluated during the follicular phase of their cycles. The PSG data were
recorded during a single night in a controlled hospital environment, with sleep
stages scored every 30 seconds. This carefully curated data set enables robust
estimation of the underlying effects of CFS and CFS+FM in early to mid-adulthood
women.

\textbf{Validation Cohorts.} The \textit{Bern Sleep–Wake Registry} (BSWR) from
the University Hospital Bern and the open-access \textit{Sleep Heart Health
Study} (SHHS) are clinical and general-population data sets used to assess the
robustness and validate the next-stage predictions of our developed model. To
ensure demographic alignment with the primary cohort, subsets of 834 and 1227
women aged 20–60 were selected from the BSWR and baseline-SHHS (SHHS1),
respectively. The BSWR challenged the model's predictive capabilities with a
population of sleep-disordered subjects, while SHHS1 assessed it in a general
population.

To ensure consistency across analyses, sleep-scoring in all data sets was
standardised to five sleep-wake stages following the AASM guidelines
\cite{berry2017aasm}: W = Wake, R = Rapid-eye-movement sleep, and (N1, N2, N3)
non-R sleep-states.

\textbf{Preprocessing.} Having a controlled homogeneous study population (women
of the same age), we considered Health Status (HS): H, CFS, CFS+FM, as the only
demographic variable. When modelling sleep dynamics, we ignored the PSG
recordings before the first non-W stage. We identified continuous bouts (runs)
of each stage---denoted $S_t$, indexed by $t$---and recorded their durations
($D_t$). This reduced the original 44,581 sleep-stage epochs to 7,254 bouts. For
each bout, we also recorded the time-since-sleep-onset ($T_t$, TSSO) and
cumulative characteristics ($C_t$) monitoring either sleep-time (CST=N1+N2+N3+R)
or restorative-sleep-time (CRST=N3+R). To utilise the existing Bayesian
inference implementation~\cite{scutari2021BNinR}, we discretised the TSSO into
five 90-minute categories ($<$90, 90-180,..., $>$360) of expected sleep cycles
and split $C_t$ and $D_t$ variables into four groups based on (25, 50,
75)\%-quantiles, with the possible additional class of 0, if present in the
corresponding variable. The validation cohorts underwent the same preprocessing,
yielding 113,071 and 150,296 bouts, respectively.

\subsection{Bayesian Networks to Capture Sleep Stage Dynamics}

A Bayesian Network (BN) is a statistical framework that encodes probabilistic
relationships between variables and can represent cause-effect relationships
under additional causal assumptions~\cite{scutari2021BNinR}. These relationships
can be learned from data (structure learning), defined by experts (incorporating
domain knowledge), or by combining both approaches. Represented as a Directed
Acyclic Graph (DAG), BNs offer several advantages, including reduced parameter
complexity and interpretable predictions—a critical requirement in the
healthcare field.

A compelling feature of BNs is their ability to fix specific nodes (variables)
at desired levels, such as the health status (HS) to H or CFS, representing what
is referred to as an \textit{intervention}. This enables do-calculus and the
simulation of causal
counterfactuals~\cite{pearl2009causality,bareinboim2016causal}, addressing
what-if questions such as ours of \textit{how sleep patterns change if a healthy
individual were to develop CFS or CFS+FM}. Dynamic BNs (DBNs) extend the
approach to temporal processes by incorporating dependencies across time,
including lagged features. This makes them particularly suited for modelling
sleep transitions.

\textbf{Experimental Setup.} Rather than relying only on data-driven structure
learning algorithms, we predefined dependencies using expert knowledge, as shown
in Fig.~1. This included mandatory (solid) edges encoding the impact of all
previous stages on the following ones (in red) and the impact of HS on $S, C, D$
(in green). The possible (dashed) impact of TSSO ($T$) on $S, C, D$ (in yellow)
was also considered. Further, we hypothesised that transitions in $S$ might be
better explained by considering cumulative sleep variables $C$ (in orange),
naturally depending on $T$. Existing work suggested that including
stage-duration $D$, depending on $S$ (in red), might boost next-stage
predictions (in blue)~\cite{yetton2018quantifying}.

To systematically evaluate each variable's inclusion and identify the optimal
structure, including BN lag/order (0–4), we fitted the BN for each possible
combination of non-mandatory nodes and their associated dependencies (edges),
and used linear regression to associate BN performance (for next-stage and HS
prediction) with indicators of each variable's inclusion. Clinically, beyond
quantifying the effects of CFS and CFS+FM (HS-node), this allowed us to test
whether the cumulative sleep (CST, CRST) better explains sleep dynamics than
TSSO. Despite the well-known effect of TSSO on sleep macro-architecture (e.g.,
higher R\% in the second half of the night), its influence on dynamics remains
inconclusive \cite{yetton2018quantifying}.

\begin{figure}
  \centering
  \includegraphics[width=0.5\textwidth]{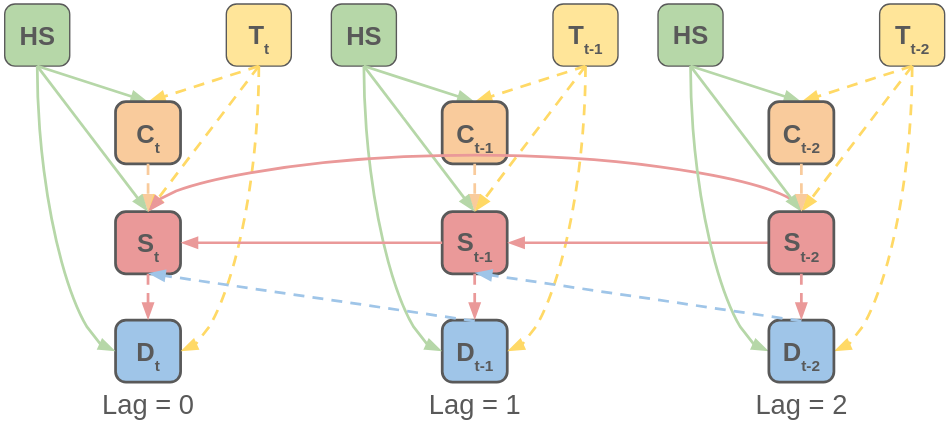}
  \caption{Illustration of the full-structure Bayesian Network of lag = 2. HS =
    health status (healthy, CFS, or CFS + FM), $T_t$ = time since the sleep
    onset, $C_t$ = cumulative sleep, $S_t$ = sleep-stage, $D_t$ = duration of
    sleep-stage, chronologically indexed by $t$.}
  \label{fig:BN_illustration}
\end{figure}

\section{Results}

\subsection{Descriptive Statistics}

\textbf{Traditional sleep variables} of the primary dataset are described in
detail in the original work, which also reports their Tukey-Kramer multiple
comparisons concerning the HS~\cite{kishi2011sleep}. The significant differences
were identified for the total sleep time [mins] (H > CFS), N1 and N2 [mins] (H >
CFS+FM), N3 [mins] (CFS+FM > H), and REM [mins] (H > CFS), c.f., Table 2 in
\cite{kishi2011sleep}.

\textbf{Occurrence of stage-specific bouts and their duration} is presented in
Table~\ref{tab:bouts}. H experienced significantly more R-bouts than both CFS
conditions and more N1-bouts than CFS+FM. In addition, CFS+FM exhibit more
N3-bouts than CFS only. The subject-aggregated means of stage-specific bout
durations did not exhibit significant differences across HS.

\begin{table}
  \centering
  \scriptsize
  \caption{Mean (SD) bout statistics for H, CFS, and CFS+FM subjects. Bouts
    indicate the average number of stage runs experienced, and Duration indicates their
    length in minutes. Significant pairwise comparisons according to the
    Tukey-Kramer procedure are marked with * and ** for p-value < (0.05 and
    0.01), respectively.}
  \begin{tabular}[t]{l|l|rrr|r}
    \textbf{Stage} & \textbf{Characteristic} & \textbf{H} & \textbf{CFS} & \textbf{CFS+FM} & \textbf{Significant Pairs}\\
    \hline
    W & Bouts & 22.5 (6.4) & 22.5 (8.5) & 19.7 (5.6) & -\\
    & Duration & 2.4 (1.5) & 3.6 (3) & 3.2 (2.1) & -\\
    \hline
    N1 & Bouts & 44.6 (17.2) & 37.5 (17.7) & 29.8 (8.9) & H - (CFS+FM)*\\
     & Duration & 1 (0.2) & 1 (0.2) & 0.9 (0.2) & -\\
    \hline
    N2 & Bouts & 50.2 (16) & 43.1 (11.9) & 52.7 (14.8) & -\\
     & Duration & 5 (2) & 5.1 (1.7) & 4.1 (2.1) & -\\
    \hline
    N3 & Bouts & 18.4 (10.2) & 15.6 (6.6) & 25.1 (12) & CFS - (CFS+FM)*\\
     & Duration & 2.2 (2.2) & 2.8 (1.9) & 3.3 (2) & -\\
    \hline
    R & Bouts & 13.4 (7) & 6.5 (4.8) & 8.2 (3.8) & H - CFS**; H - (CFS+FM)*\\
     & Duration & 8.2 (5.8) & 12.6 (8.5) & 10.8 (7.9) & -\\
    \hline
  \end{tabular}
  \label{tab:bouts}
\end{table}

\begin{table}
  \centering
  \scriptsize
  \caption{The impact of BN-included variables on the performance metrics.
    Significant variable associations and model explanations based on F-test are
    highlighted as p-value < \highlight{0.05}{blue!5},
    \highlight{0.01}{blue!10}, and \highlight{0.001}{blue!20}, respectively. The
    $R^2_{\text{adjusted}}$ (not tested) and F-statistic refer to regression
    models evaluating the systematic impact of included variables on the
    performance metric across different BN-settings and not to any specific BN.}
  \centering
  \begin{tabular}{lrrr}
    \hline \hline
    \textbf{Variable} & \textbf{Accuracy} [$S_t$] & \textbf{F1-score} [$S_t$] & \textbf{AUROC} [HS] \\
    \hline
    lag = 0 & \cellcolor{blue!20}44.00 & \cellcolor{blue!20}50.84 & \cellcolor{blue!20}71.59 \\
    lag = 1 & \cellcolor{blue!20}68.03 & \cellcolor{blue!20}72.75 & \cellcolor{blue!20}74.13 \\
    lag = 2 & \cellcolor{blue!20}72.08 & \cellcolor{blue!20}73.07 & \cellcolor{blue!20}74.29 \\
    lag = 3 & \cellcolor{blue!20}68.91 & \cellcolor{blue!20}70.05 & \cellcolor{blue!20}75.63 \\
    lag = 4 & \cellcolor{blue!20}64.78 & \cellcolor{blue!20}65.94 & \cellcolor{blue!20}75.85 \\
    TSSO    & \cellcolor{blue!20}-4.14  & \cellcolor{blue!20}-4.62 & \cellcolor{blue!20}-4.14 \\
    Stage-Duration & \cellcolor{blue!10}-3.29  & \cellcolor{blue!20}-3.87 & \cellcolor{blue!20}2.45 \\
    CST & -1.53      & -1.65      & \cellcolor{blue!20}-3.98 \\
    CRST & \cellcolor{blue!20}-12.03 & \cellcolor{blue!20}-12.56 & \cellcolor{blue!20}-8.42 \\
    \hline \hline
    Model's F(9, 51) & \cellcolor{blue!20}1252.09 & \cellcolor{blue!20}2323.33 & \cellcolor{blue!20}4715.47 \\ \hline
    Model's $R^2_{\text{adjusted}}$ & 0.995 & 0.997 & 0.999 \\
  \end{tabular}
  \label{tab:CV}
\end{table}

\begin{figure}
  \centering
  \includegraphics[scale=0.3]{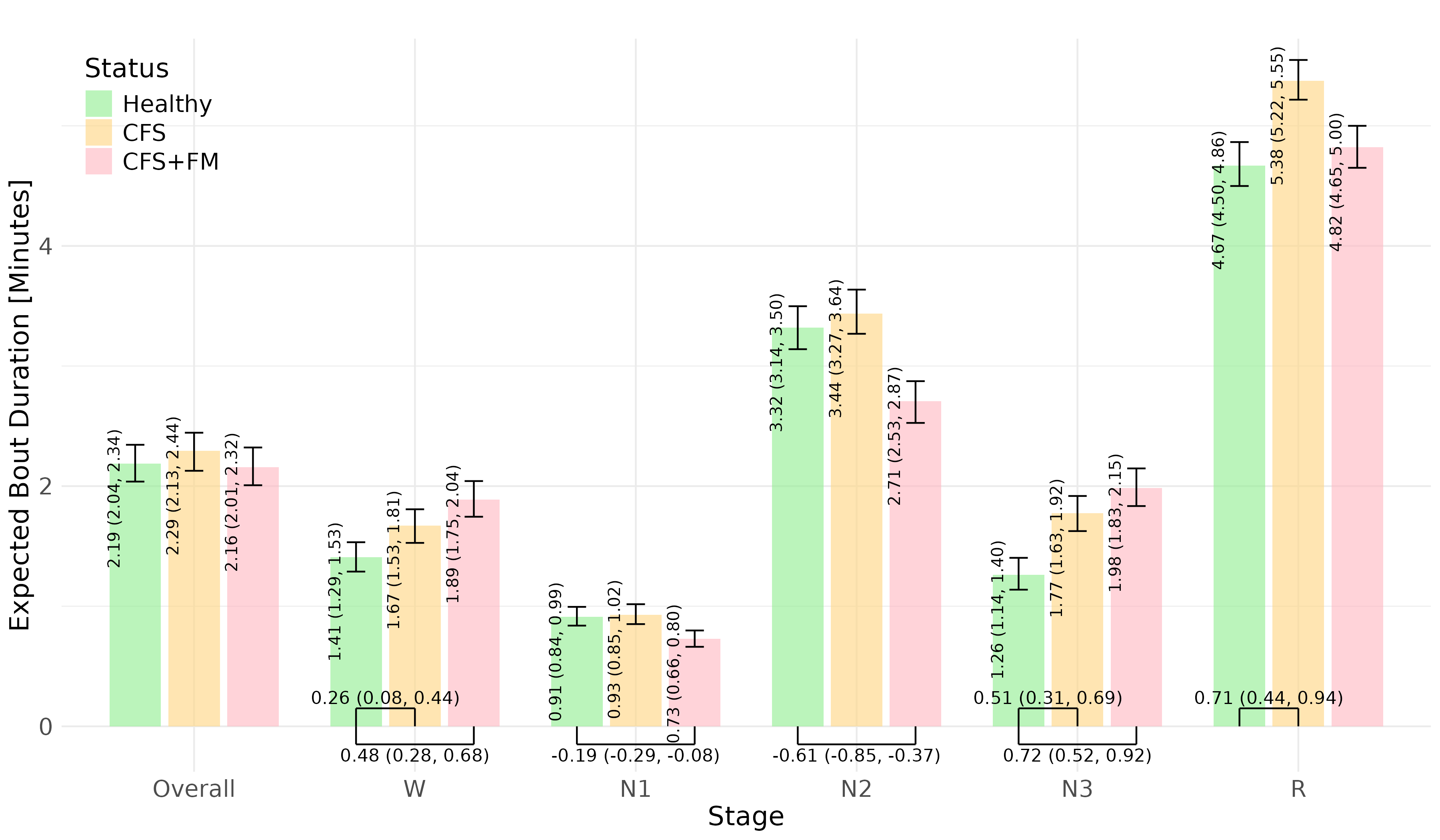}
  \caption{Expected durations of sleep-stage-bouts for H, CFS, and CFS+FM
    groups, presented with 95\% CI as vertical error bars. Horizontal brackets
    indicate significant differences between conditions, accompanied by their
    estimates and 95\% CI.}
  \label{fig:durations}
\end{figure}

\subsection{Structure Identification}

The structure of BN was selected based on the computational experiment described
above, testing all expertly predefined node combinations from
Fig.~\ref{fig:BN_illustration} under restricted settings of temporal ordering
and HS being the underlying cause of transitions. This evaluation used
HS-balanced 3-fold cross-validation (CV) with subject-wise splits, allowing
performance quantification for each variable combination while ensuring a
reasonable number of subjects were included in the testing fold. The performance
metrics used included: \textbf{next-stage accuracy} and \textbf{F1-score}, and
\textbf{average AUROC} (of AUROCs specific to H, CFS, and CFS+FM). The results
are summarised in Table~\ref{tab:CV}. Based on next-stage performance metrics,
we identified lag=2 as optimal, confirming~\cite{yetton2018quantifying}, as both
accuracy and F1-score were the highest and appear to decrease with larger lags.
The AUROC, indicating the capability to identify HS, was up to 1.56\% better for
higher lags, but their consideration would lead to an expected decrease of up to
7.3\% in accuracy/F1-score. All TSSO, CST, and CRST yielded a systematic
decrease in all performance metrics. This may suggest that sleep dynamics and HS
identification are either unrelated to these variables or that the BN was under
their inclusion over-parametrised, as the number of parameters to predict the
$S_t$ just from $S_{t-1}, S_{t-2},$ HS involves 75 = 5$\times$5$\times$3
parameters which scale by 4-5 with the inclusion of every additional (TSSO, CST,
CRST) variable. Based on that, we chose the BN of lag = 2 with included stage
durations as the final model to demonstrate the CFS and CFS+FM effects. Despite
slightly reduced next-stage predictive accuracy due to duration inclusion, this
model seems to significantly enhance the identification of HS. Our evaluations
tried to find a compromise between the best performance in the next stage and
diagnosis identifications.

\subsection{Performance and Generalization}

The final BN (lag = 2, including stage durations) achieved 70.61 (1.9)\% and
69.2 (2.7)\% in mean (SD) on-subject next-stage accuracy and F1-score, and the
HS AUROC of 75.36 (8.3)\%. For each subject, we estimated HS probabilities by
averaging posterior queries over all triplets of sleep stages and durations.

To further test the robustness of the final BN to capture sleep dynamics, we
evaluated its predictive accuracy on BSWR and SHHS1. Despite training on a small
sample of 52 strictly controlled subjects, BN achieved 69.78 (7.25)\% and 60.1
(11.62)\% in mean (SD) on-subject accuracy, 70.94 (9.1)\% and 59.83 (11.56)\% in
on-subject F1-score, on BSWR and SHHS1, respectively. Considering that both test
data sets represent out-of-domain samples from general and clinical cohorts,
respectively, with considerable domain shifts, these results suggest the
particular robustness of our BN. In contrast, similar work reported 62.2\%
testing accuracy (corresponding to in-domain cross-validation assessment) on a
broad sample of 3,202 PSG recordings with excluded
sleep-disorders~\cite{yetton2018quantifying}.

\subsection{Effects of CFS and CFS+FM via Interventions}

We evaluated three interventions by fixing the HS node of our final BN to H,
CFS and CFS+FM levels, allowing sampling from arbitrary nodes under specified
conditions. Assuming no hidden confounding, which is reasonable in our strictly
controlled cohort, comparing samples for CFS-vs-H and (CFS+FM)-vs-H enables
estimating the causal effects of the two conditions. Arbitrary 95\% credible
intervals (CI) were constructed by generating 1,000$\times$1,000 samples and
calculating median (= estimate) and (2.5, 97.5)\%-quantiles (= CI-bounds).

\textbf{Bouts Duration:} Fig.~\ref{fig:durations} presents BN-based CIs for
expected stage durations. Discretised $D_t$ levels were represented by
mid-points and multiplied by the obtained samples. Both CFS and CFS+FM exhibit
prolonged W and N3 durations, indicating reduced sleep efficiency and increased
physically-restorative drive. CFS additionally exhibits extended R stages,
linked to cognitive restoration, despite fewer R bouts. In contrast, CFS+FM
shows shorter N1 durations, likely compensating for increased W and N3. Notably,
CFS does not display reduced durations in any stage, suggesting compact sleep
despite decreased efficiency.

\begin{figure}
  \centering
  \includegraphics[width=\textwidth]{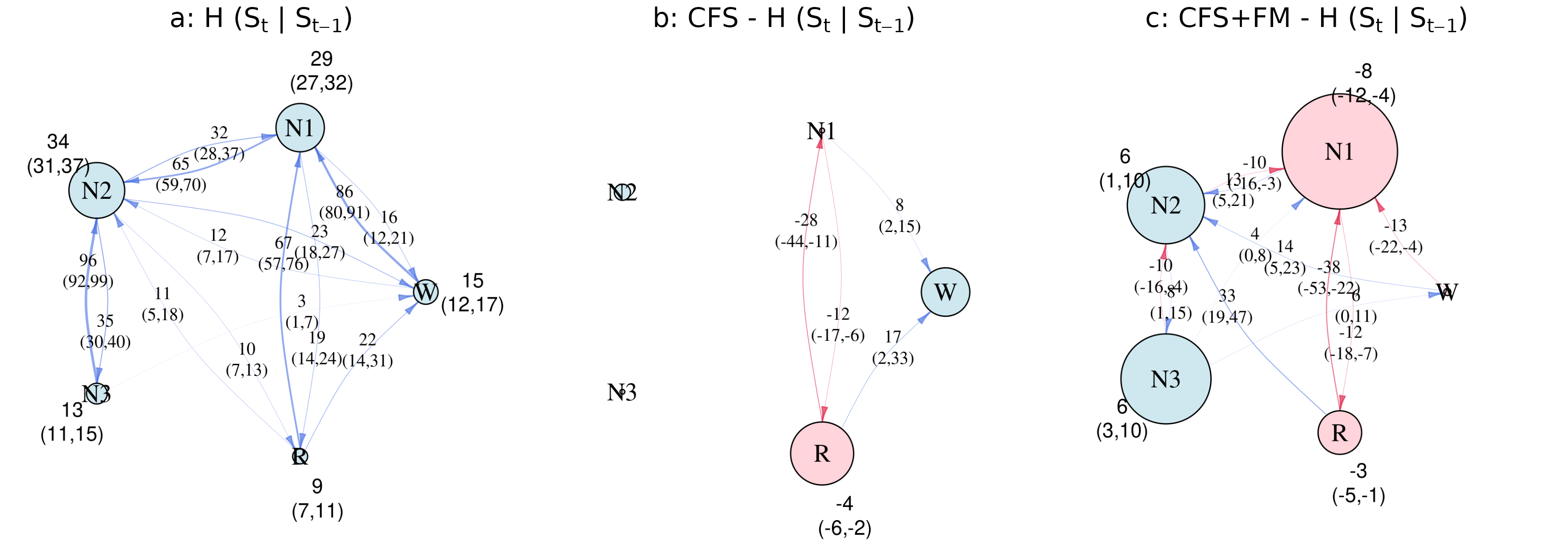}
  \caption{Lag-1 sleep-stage transition dynamics for Healthy (H), Chronic
    Fatigue Syndrome (CFS), and CFS with Fibromyalgia (CFS+FM). Panel (a)
    illustrates the expected transitions for H, with node sizes proportional to
    the prevalence of $S_{t-1}$ stages and edges indicating transition $(S_t
    \mid S_{t-1})$ probabilities. Panels (b) and (c) depict the differences in
    stage prevalence and transition probabilities due to CFS and CFS+FM, in
    comparison to H, respectively. Positive and negative values are shown in
    blue and red, respectively, and significant alterations are annotated with
    their estimates and 95\% credible intervals.}
  \label{fig:lag1trans}
\end{figure}

\textbf{First-order transitions} $(S_t \mid S_{t-1})$ expected for H are shown
in Fig.~\ref{fig:lag1trans}.a and the CFS and CFS+FM effects in
Fig.~\ref{fig:lag1trans}.(b-c). The effects were quantified without conditioning
on any particular stage and describe the overall sleep dynamics. Below, we write
in \textbf{bold} alterations by at least 10\%. CFS showed reduced R\% and
increased N1\ra W, \textbf{R\ra W} that were compensated
by decreased \textbf{N1$\rightleftarrows$R}. The changes were more pronounced in
CFS+FM, which showed increased (N2, N3)\% and decreased (N1, R)\%. Further,
CFS+FM exhibited significantly increased (W, \textbf{N1},
\textbf{R})\ra \textbf{N2}, N2\ra N3, N3\ra
(W, N1), and decreased \textbf{(W, N2, R)\ra N1},
\textbf{N1\ra R}, and \textbf{N3\ra N2}. Our findings
confirm all alterations found by \cite{kishi2011sleep} in their Fig.~1. We
additionally identified increased N1\ra W
(c.f.,~\cite{kishi2008chronic_fatique}) in CFS (compensation for decreased
N1\ra R) and disruptions in N2 for CFS+FM~\cite{kishi2011sleep}.

\begin{figure}
  \centering
  \includegraphics[width = 0.9\textwidth]{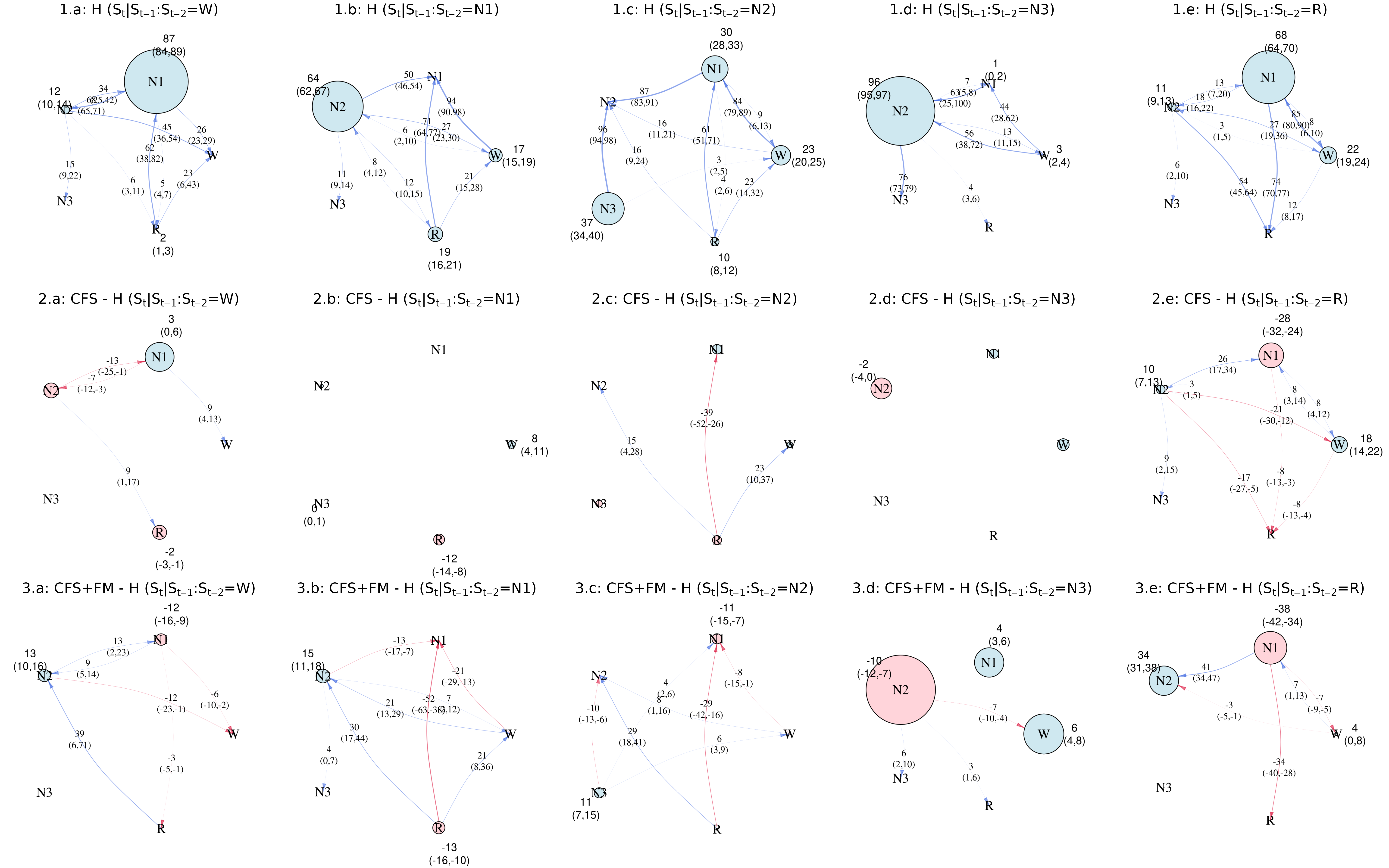}
  \caption{Lag-2 sleep-stage transition dynamics $(S_t \mid S_{t-1}, S_{t-2})$
    for Healthy (H), Chronic Fatigue Syndrome (CFS), and CFS with Fibromyalgia
    (CFS+FM). The first row shows the expected dynamics for H, while rows two and
    3 display their changes due to CFS and CFS+FM, in comparison to H,
    respectively. Node sizes represent $S_{t-1}$ prevalence (or its difference),
    and edges illustrate transition probabilities $(S_t \mid S_{t-1})$ or their
    differences, both conditioned on $S_{t-2}$. Positive and negative values are
    shown in blue and red, respectively, and significant alterations are
    annotated with estimates and 95\% credible intervals.}
  \label{fig:lag2_transitions}
\end{figure}

\textbf{Second-order transitions} $(S_t \mid S_{t-1}, S_{t-2})$ in
Fig.~\ref{fig:lag2_transitions} provide deeper insights into sleep dynamics.
transitions for H, while rows 2 and 3 depict the effects of CFS and CFS+FM. Each
column (a--e) corresponds to a different starting stage $S_{t-2}$. In some cases,
the alterations are only in $S_{t-1}$ (nodes), or follow-up transitions
($S_{t-1}$\ra $S_t$, edges), both conditioned on $S_{t-2}$ and
extending the unconditioned first-order results from Fig.~\ref{fig:lag1trans}.

In \textit{\textbf{CFS}}, key disruptions included increased
\textbf{R\ra W} (with subsequent increases in N1 and decreases in R)
and \textbf{R\ra N2} (followed by increased \textbf{N1}, N3, and
decreased \textbf{W}, \textbf{R}), along with reduced \textbf{R\ra
N1}. These patterns suggest an impaired ability to achieve or maintain
restorative R sleep, compensated by non-restorative transitions within light
sleep (N1, N2). Additionally, W\ra R, common in healthy individuals
during the second half of the night, was decreased and supplemented by
W\ra N1. More frequent N1\ra W, at the expense of
\textbf{N1\ra R}, further contributed to reduced sleep-efficiency
and increased fragmentation.

In \textit{\textbf{CFS+FM}}, disruptions included increased R\ra W
(with subsequent increases in N1 and decreases in N2) and
\textbf{R\ra N2}, along with reduced \textbf{R\ra N1}
(followed by decreased W and \textbf{R}, and increased \textbf{N2}).
Particularly increased \textbf{N2\ra N3} (followed by increased
transitions to W and N1, and reduced to \textbf{N2}), reflecting a compensatory
drive for deep sleep (N3) likely linked to FM's restorative needs, while also
indicating difficulty maintaining smooth sleep cycling. Reduced
\textbf{W\ra N1} and increased \textbf{W\ra N2} suggest
a shift towards intermediate sleep stages at the expense of lighter sleep,
possibly as a response to pain-related disruptions. Increased awakenings from N3
(compensated by reduced \textbf{N3\ra N2}) and from R further
destabilised transitions between restorative and lighter stages, amplifying
sleep fragmentation and reducing efficiency. These findings align with FM's
symptomatology, where widespread pain increases the need for deep sleep (N3) but
disrupts restorative sleep transitions, highlighting the need for tailored
treatments to improve both sleep and pain management.

\section{Discussion}

In this study, we constructed a Bayesian Network (BN) to quantify the effects of
Chronic Fatigue Syndrome (CFS) and its interaction with Fibromyalgia (FM) on
sleep dynamics. Using a strictly controlled dataset~\cite{kishi2011sleep}, we
confirmed that second-order transitions  $(S_t \mid S_{t-1}, S_{t-2})$ optimally
describe sleep patterns, extending findings from non-clinical
populations~\cite{yetton2018quantifying}. Despite a relatively small dataset of
7,254 bouts from 52 subjects, our BN achieved robust next-stage predictions with
in-domain (out-of-domain) accuracies of 70.6\% (60.1–69.8\%), respectively. This
capability enabled the successful differentiation of healthy (H), CFS, and
CFS+FM groups (AUROC: 75.4\%), showcasing sleep dynamics' potential for
diagnostics. Based on that, we used interventions to quantify the effects of CFS
and CFS+FM compared to H on different aspects of sleep dynamics.

Both conditions exhibited prolonged wakefulness (W) and N3 stages, reflecting
reduced sleep efficiency (aligning with insomnia symptoms in
CFS~\cite{unger2004sleep}) and increased physical restoration needs,
particularly pronounced in CFS+FM. Additionally, CFS showed extended R durations
related to an increased sympathetic activity and a higher need for cognitive
restoration, while CFS+FM demonstrated reduced durations of N1 and N2.
Interestingly, the duration of any stage did not decrease in CFS, suggesting
that their sleep, despite reduced efficiency, may remain relatively compact.%
, echoing findings on preserved sleep architecture despite
fatigue~\cite{natelson2019myalgic}.

First-order transitions confirmed all previous findings~\cite{kishi2011sleep},
and - thanks to the joint estimation of transition probabilities in our BN (as
opposed to the pairwise comparisons in \cite{kishi2011sleep}), revealed three
additional compensatory transitions. CFS is marked by frequent and prolonged
awakenings from the N1 and R stages, disrupting "healthy" oscillations between
them. This suggests reduced sleep efficiency at the expense of R sleep,
potentially contributing to fatigue from both, insufficient sleep quantity and
inadequate autonomic or cognitive restoration. In contrast, CFS+FM is
characterised by awakenings from deep N3 sleep, into which they tend to
transition more frequently. FM, associated with physical pain and
discomfort~\cite{galvez2020diagnostic}, appears to drive both the increased N3
duration and the pressure to transition to N2 instead of N3 across stages.

The second-order transitions provided a novel and detailed perspective on sleep
alterations. Both conditions exhibited increased transitions into W,
particularly from R, reflecting reduced sleep efficiency. For CFS, fewer
alterations were observed, consistent with their longer bouts. The results
highlighted CFS-specific patterns of awakenings from N1 and R, difficulties
maintaining R (due to transitions into N2), and challenges achieving R. These
disruptions may represent the patients' common complaint of "unrefreshing
sleep", either as a cause or a consequence of fatigue, as commonly reported in
CFS. In contrast, CFS+FM showed more widespread alterations, including frequent
awakenings from both R and N3, coupled with a marked compensatory drive to
achieve and sustain N3, likely driven by the physical symptoms of FM.

\paragraph{Conclusions:}

Our study confirms that sleep transitions are best described as a second-order
process, even in diseased clinical subjects. Using a strictly controlled cohort
of young-to-middle-aged women, we identified the effects of CFS and CFS+FM on
altered sleep and its dynamics, supporting their clinical differentiation.
These findings highlight the potential of sleep dynamics as a non-invasive
diagnostic tool and may suggest differing therapeutic needs tailored to the
unique sleep disruptions observed in these conditions. Our findings should not
be directly generalised to males and older subjects, as our study population did
not include them, necessitating further evaluations in these groups.


\end{document}